# On the processing of anatase La-doped TiO$_2$ nanopowders: structural phase transition and surface defect engineering

M. P. Ferreira, and H. B. de Carvalho*

$^a$ Universidade Federal de Alfenas – UNIFAL-MG, 37130-001 Alfenas, MG, Brazil

*Corresponding Authors*:
* hugo.carvalho@unifal-mg.edu.br

**ABSTRACT:** The efforts in the optimization of the physical and chemical properties related to the photocatalytic efficiency of TiO$_2$ nanoparticles relies mainly on defect engineering strategies, been the structural and the doping the most common ones. The doping of the anatase TiO$_2$ structure with lanthanides elements potentially can turns the TiO$_2$ optical absorption into the visible range increasing drastically its photocatalytic efficiency. Here we show that the processing of TiO$_2$ nanoparticles in oxygen atmosphere and the La doping stabilizes the anatase phase. The La incorporation takes place mainly in the surface of the nanoparticles leading to small nanoparticles as compared to the undoped TiO$_2$. In spite of the effective La doping, the decrease of the energy bandgap is compensated by a quantum confinement effect related to the decrease in the nanoparticle size. However, the location of the La$^{3+}$ ions on the surface of the nanoparticles induces a local concentration of oxygen vacancies that can increase the quantum efficiency of the system leading to higher photocatalytic activity.

## 1 INTRODUCTION

Among the semiconductors oxides presenting photocatalytic properties, nanostructured

TiO$_2$ has been the most studied in recent decades [1, 2]. However, its photocatalytic efficiency is

limited mainly by two main factors: its optical absorption in the UV region of the electromagnetic

spectrum and its relatively low quantum efficiency, high recombination rate of the photogenerated

$e^-/h^+$ pair [3]. The metastable anatase phase of the TiO$_2$ present superior photocatalytic efficiency

as compared to the rutile phase due to its relatively higher carrier mobility, what favor the separation of the photogenerated $e^-/h^+$ pair inhibiting its recombination [4]. The proposed strategies to extend the TiO$_2$ absorption into the visible spectrum include: surface modification and sensitization with dyes, deposition with noble and non-noble metals, heterostructuring with materials with small bandgap, and defect engineering [1].

Specifically, the introduction of defects and elements (doping) into the $TiO_2$ structure aims to introduce electronic states inside the bandgap, shifting the $TiO_2$ absorption to the visible part of the electromagnetic spectrum [5-7]. Bae *et al*. demonstrated that $Co^{2+}$ and $Cr^{3+}$ doping of $TiO_2$ reduces the anatase $TiO_2$ bandgap from 3.2 eV to 1.82 and 2.31 eV, respectively [8]. On the literature one can find reports of the $TiO_2$ doping by earth metals ($Ca^{2+}$, $Sr^{2+}$, $Ba^{2+}$) [9], $Fe^{3+}$ [10, 11], $Mo^{5+}$ [12, 13], and, in particular, elements of the lanthanide family ($La^{3+}$, $Ce^{3+}$, $Pr^{3+}$, $Nd^{3+}$, $Sm^{3+}$, $Gd^{3+}$ and $Er^{3+}$) [14]. The $4f$ orbitals of the lanthanides lie within the bandgap close to the $TiO_2$ conduction band, thus the $TiO_2$ doping with these elements can effectively reduce its bandgap [15]. Lanthanides also have a large crystalline radius compared to $Ti^{4+}$, in such a way that their insertion in the $TiO_2$ structure promotes distortions that stabilize the anatase phase [16-18]. The doping of $TiO_2$ with elements of the lanthanide family (with oxidation state +3) also favors the formation of oxygen vacancies ($V_O$) via charge compensation processes. There are experimental reports that point out that specific defects, such as $V_O$, can contribute to increase the photocatalytic efficiency [19, 20], especially when located on the surface of the nanoparticle [21, 22]. Lanthanides are known to easily form Lewis complexes with bases through the interaction of these functional groups with their $f$ orbitals [23, 24]. In this way, lanthanide elements incorporated into the surface of the $TiO_2$ nanoparticle can potentially bind more easily to target molecules to be degraded, increasing their adsorption (concentration) on the surface of the nanoparticle and, thus, increasing the photocatalytic efficiency of the material [23].

In this context, the present work presents a detailed study of the main characteristics and properties related to the preparation of nanostructured $TiO_2$ powders in its anatase phase, pointing out the main parameters necessary for the stabilization of the desired phase. The chosen technique for preparing the samples was the polymeric precursor method, as it provides samples in the nanometer scale in a simple way, it is cheap and allows a great homogeneity of the precursor elements, which disfavors the formation of unwanted secondary phases [25]. From the obtained results, we prepared La-doped anatase TiO₂ samples and evaluated its structural and physical properties regarding its application in photocatalytic systems.

## 2 EXPERIMENTAL METHODS

Nanostructured $TiO_2$ and $La:TiO_2$ samples were synthetized via the polymeric precursor method. The metal sources in the synthesis were Titanium tetraisopropoxide, $Ti[OCH(CH3)2]4$ (Sigma-Aldrich, 97%), and Lanthanum oxides, $La_2O_3$ (Sigma-Aldrich, 99%). Citric acid (CA) and ethylene glycol (EG) were used to complex the metallic cations. First, titanium tetraisopropoxide was dissolved in an CA aqueous solution at 100º C under constant stirring to obtain the Ti citrate. The La citrate was obtained by the dissolution of the $La_2O_3$ in aqueous nitric acid ($HNO_3$) solution. After a complete dissolution, CA was added. Both citrates were than mixed at 60 °C under constant stirring. After the formation of a completely homogenous solution the temperature was raised up to ~110 °C and EG was added to promote a polymerization by polyesterification. This solution was kept under stirring for ~ 15 minutes. Then, the temperature was decreased to 90 °C and kept at continuous stirring until a viscous resin was formed. The used CA/metal molar ratio and the CA/EG mass ratio were 8:1 and 60:40, respectively. The polymeric precursor resin was heat treated at 300° C for 2 h at a heating rate of 2° C/min. The obtained solid (puff) was grounded in an agate mortar. Finally, the powder was heat treated in different temperatures and atmospheres in order to determine the best parameters of synthesis.

Thermal analyses, thermogravimetry (TG) and differential scanning calorimetry (DSC), were performed in TA Instruments Q600 with a continuous flux of air (100 ml/min.) in alumina crucibles in the temperature range of 20 to 1000 °C (heating rate of 10 °C/min.). The structural properties were investigated by X-ray diffraction (XRD) recorded in the range of 2θ = 20°–80° with steps of 0.02° at 10 s/step by using Cu-Ka radiation (λ = 1.542 Å) of a Rigaku Ultima IV diffractometer. Raman measurements were carried out at room temperature on a modular spectrometer consisting of an Olympus B-X41 microscope and a Horiba iHR550 monochromator at the backscattered photon detection geometry. The spectra were collected at several different points for each sample in order to enhance the statistical analysis. The excitation was performed with a 532 nm wavelength laser and power of 1 mW over the sample surface. The morphology and structure were characterized by using a Tecnai G2-20 SuperTwin FEI high-resolution transmission electron microscopy (HRTEM), operating at 200 kV. Elemental analyses were performed in a JEOL-JSM 6510 equipped with Oxford X-MAX 80 energy dispersive X-ray spectrometer. Photoreflectance measurements were performed in the range of 200−800 nm in a Lambda 1050 Perkin Elmer spectrophotometer.

### 3   RESULTS AND DISCUSSION

Figure 1 presents the (a) TG and (b) DSC curves obtained after the heat treatment of the $TiO_2$ puff. In Figure 2, we have the diffractograms and Raman spectra obtained by heat treating the puff at the temperatures indicated in Figure 1. Here the diffractograms and spectra are normalized by the (101) diffraction peak of the anatase phase and (110) of the rutile phase and by the corresponding most intense vibrational mode of the anatase and rutile $TiO_2$ phases. The heat treatment of the samples was carried out in a muffle furnace in air atmosphere. The association between the thermal analyzes, the XRD diffractograms, and Raman spectra allowed us to determine the $TiO_2$ chemical kinetics of synthesis. We initially observed an endothermic peak around 100 °C corresponding to a small loss of mass of the order of 5 % associated with evaporation of residual water. In the sequence there are two successive exothermic peaks, one at approximately 390 °C and the other at 475 °C, accompanied by an abrupt loss of mass of about 87%. The first exothermic peak at approximately 390 °C is associated with the burning of organic material and crystallization in the anatase phase [26, 27]. Firstly, we could infer that the second peak at 475 °C was related to the anatase to rutile phase transition [27], however, we did not observe an abrupt and defined structural transition in the temperature range of this exothermic peak (Figure 2). Therefore, we associate this peak mainly with the combustion of residual organic material present in the sample. We can see in Figure 2(a) that for the heat treatment at 300 °C, corresponding to the puff preparation temperature, the material is completely amorphous. At 400

°C, it is obtained the desired anatase $TiO_2$. Here we call attention to the relative larger width of the diffraction peak (101) with respect to the diffractograms obtained from materials heat treated at higher temperatures. This is an evidence that the size of particles of the powder are relatively small. At 450 °C, the predominant anatase phase still predominant, but a small fraction of rutile it is observed. At 550 °C the rutile phase increases considerably, and at 650 °C the rutile became the predominant phase. The analysis of the XRD data is supported by the Raman results (Figure 2(b)).

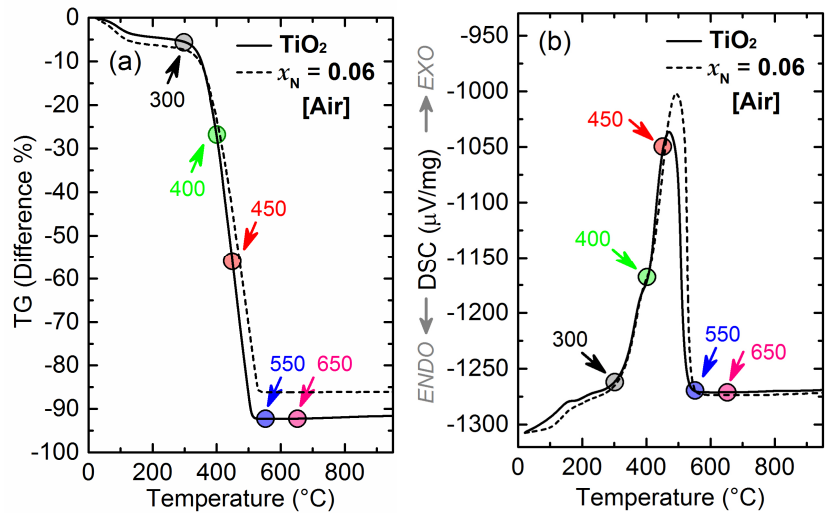

**Figure 1.** (a) Thermogravimetry (TG) and (b) differential scanning calorimetry (DSC) data of the $TiO_2$ sample and the La:$TiO_2$ sample with the La nominal content of $x_N = 0.06$.

The observation of a small amount of rutile phase at low temperatures (450 °C in Figure 2) and the broad range in temperature for the conversion of the anatase to the rutile structure reveals the metastable character of the anatase $TiO_2$ structure. In fact, one of the factors that influences the thermodynamic and kinetic process of the anatase-rutile phase transition is the particle size of the $TiO_2$ powder, when the size is in the nanometer scale the total free energy in the anatase phase is lower than in the rutile phase [28]. For larger particles, the rutile phase becomes more stable due to the increased contribution of the volume term to the free energy of the system [29]. Therefore, with increasing the temperature of the heat treatment the particle size increases due to the expected sintering process, leading to the increasing proportion of the rutile phase in the processed material. However, due to the metastable character of the anatase phase, specific local structural disorder, like clusters of point defects, would play an important role in the $TiO_2$ phase transition.

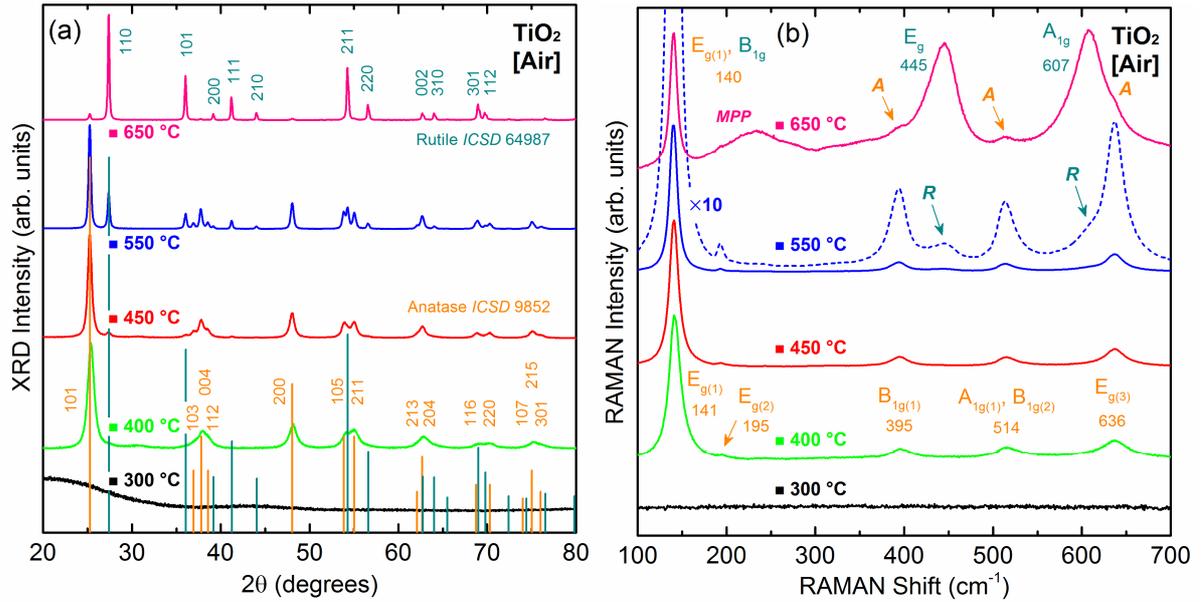

**Figure 2.** (a) XRD diffractograms and (b) Raman spectra of TiO₂ samples heat treated at 400, 450, 550 and 650 °C in a muffle furnace in an air atmosphere. For comparison, the diffractograms are normalized by the most intense diffraction peak ((101) of the anatase phase at temperatures of 400, 450, 550 °C and (110) of the rutile phase at the temperature of 650 °C), the Raman spectra are also normalized by the higher intense vibrational mode ($E_{g(1)}$ of the anatase phase and $B_{1g}$ of the rutile phase).

In order to study the dependence of the anatase-rutile phase transition on point defects, we performed the heat treatment of the samples in a tube furnace with controlled atmosphere of oxygen ($O_2$). Figure 3 presents the diffractograms and Raman spectra obtained for TiO₂ heat treated at different temperatures in a $O_2$ flow of 2.5 L/min.. Here we clearly observe that, compared to the samples heat treated in air (Figure 2), the rutile phase is substantially suppressed in the $O_2$. We also analyzed the dependence of the observation of the rutile phase on the $O_2$ flux for the heat treatment at the temperature of 500 °C. The obtained results are shown in Figure 4. Here we observe that the transition temperature, inferred from the observation of the rutile phase, is a direct function of the $O_2$ flow in our system. For the flow of 1.5 L/min. the rutile phase is present again (comparison with Figure 3), however, with an increase in the flow to 4 L/min. we did not observe significant changes in the crystallinity of the samples, given that the diffractograms (width at half height and intensity (not shown)) and Raman spectra are essentially the same as compared to the sample processed at the $O_2$ flow of 2.5 L/min.. These results lead us to infer that $V_O$ are the point defects responsible for the observed behavior. Similar results have already been reported in the literature [30, 31]. The rearrangement and transformation of the structure at the anatase-rutile phase transition is facilitated by the relaxation (decreased structural rigidity) of the oxygen sub-lattice through the presence of $V_O$ defects [32-34]. Thus, processing $TiO_2$ in atmospheres rich in oxygen favors the anatase phase, eliminating the $V_O$ and raising the temperature of the transition to the rutile phase [35, 36].

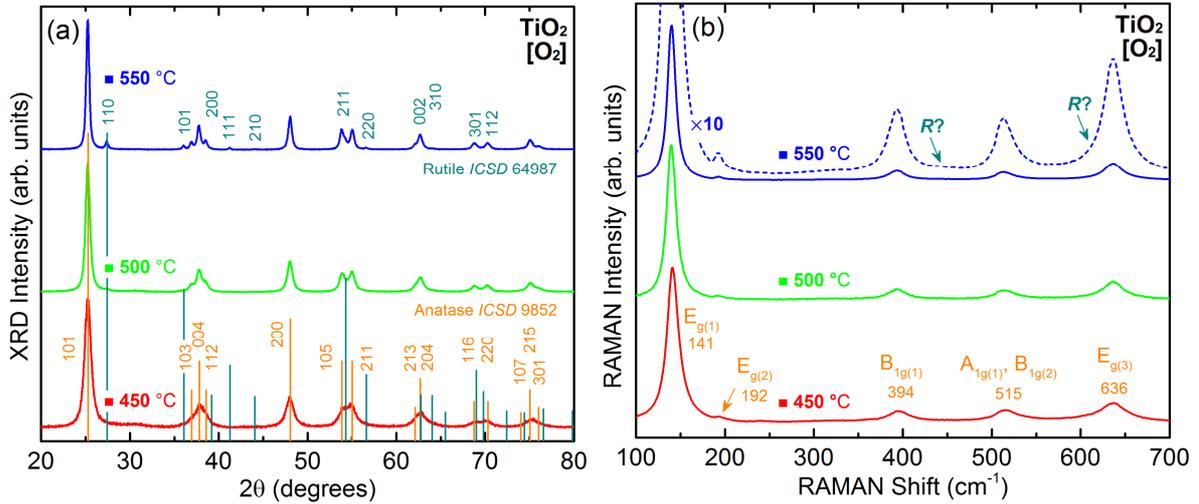

**Figure 3.** (a) XRD diffractograms and (b) Raman spectra for the $TiO_2$ samples prepared at the temperatures of 450, 500 and 550 °C in pure $O_2$ atmosphere with flow rate of 2.5 L/min. For comparison, the diffractograms are normalized by the most intense diffraction peaks ((101) of the anatase phase at temperatures of 400, 450, 550 °C and (110) of the rutile phase at a temperature of 650 °C), the Raman spectra are also normalized by the more intense vibrational mode ($E_{g(1)}$ of the anatase phase and $B_{1g}$ of the rutile phase).

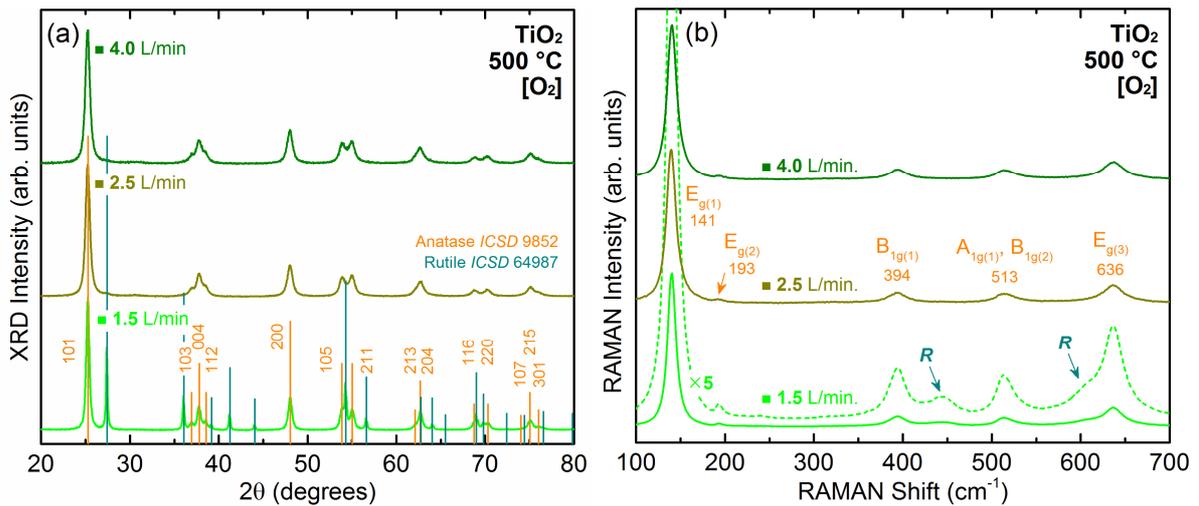

**Figure 4.** (a) XRD diffractograms and (b) Raman spectra of $TiO_2$ samples prepared at the temperature of 500 °C in $O_2$ atmosphere with flow rates of 1.5; 2.5 and 4 L/min. For comparison, the diffractograms are normalized by the most intense diffraction peak ((101) of the anatase phase), the Raman spectra are also normalized by the highest vibrational mode intense ($E_{g(1)}$ of the anatase phase).

In the sequence we proceed processing the La-doped anatase $TiO_2$ nanoparticles. Figure 5 shows the diffractograms for $La:TiO_2$ samples with the La nominal concentration of $x_N = 0.03$; 0.06 and 0.09, once considering the formation of the ternary compound $Ti_{1-x}La_xO_2$. These samples were heat treated in air in a muffle furnace at temperatures of 450 and 550 °C. The TG and DSC curves obtained for $La:TiO_2$ with $x_N = 0.06$ sample are presented also in Figure 1. It can be observed that the presence of the La does not substantially changes the thermal and chemical kinetics as compared to the obtained, and previous discussed, for the $TiO_2$ sample. The diffractograms for the $TiO_2$ sample heat treated in the same temperatures presented in Figure 2 are also shown for comparison. The diffractograms here are not normalized, so that we can evaluate the obtained crystallinity.

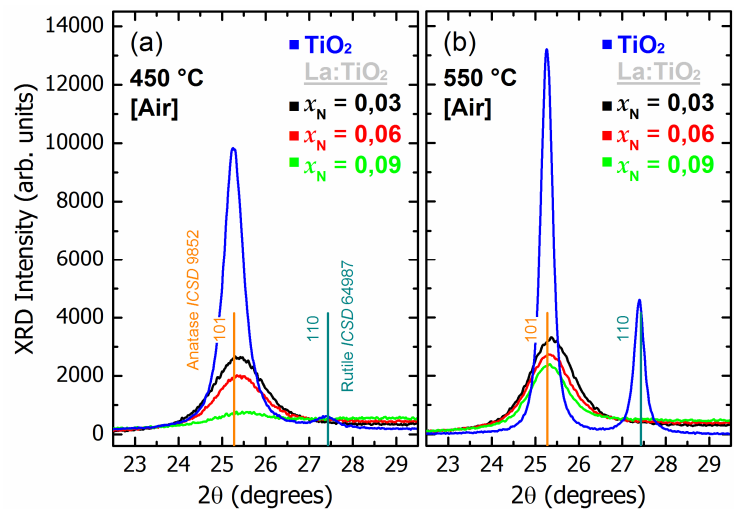

**Figure 5.** XRD diffractograms around the most intense diffraction peaks of the anatase and rutile phases for the $TiO_2$ and the $La:TiO_2$ samples with Ln nominal concentration of $x_N = 0.03$; 0.06 and 0.09 ($Ti_{1-x}La_xO_2$) heat treated at the temperatures of (a) 450 and (b) 550 °C.

We first observed for samples $TiO_2$ the presence of the rutile phase, it is also noteworthy that heat treating at 550 °C significantly leads to a greater crystallinity as compared to that for sample heat treated at 450 °C, the intensity of the peaks of diffraction increases, while widths decrease [37]. With the addition of the La, the diffraction peak associated with the rutile phase disappears. This result is indicative of the insertion of $La^{3+}$ into the anatase $TiO_2$ structure (La doping). We also call attention to the relative decrease in intensities and the widening of the diffraction peaks for the samples with La content, evidencing a significant decrease in the crystallinity for these samples and a decreasing of the size of the nanoparticles. As the dopant concentration increases, the crystallinity and the size of the nanoparticles decreases even more, for the $x_N = 0.09$ sample the parameters are relatively quite poor. The nanostructured character of the prepared powders, and the decrease of the size of the nanoparticles with the La concentration, and the effective La doping of the anatase $TiO_2$ structure was further confirmed in the subsequent analysis.

In the literature we found several experimental reports indicating that doping can either favor or disfavor the phase transition. In the case of a substitutional solid-state solution (substitutional doping), the doping atoms can enter the anatase of $TiO_2$ structure, favoring the formation of $V_O$ and, thus, promoting the anatase-rutile phase transition. Besides, when an interstitial solid-state solution is formed (interstitial doping), the insertion of an element into an interstitial site ends up imposing mechanical restrictions that inhibit the structural conformations necessary for the phase transition, thus leading to its inhibition [16-18]. Numerous elements have been used in the cationic doping of $TiO_2$, in general, considering a substitutional doping, it is observed that cations with an oxidation state lower than +4 replacing the $Ti^{4+}$ favor the anatase-rutile transition due to the induction of $V_O$ via charge compensation processes [30, 32-34, 38]. On the other hand, the replacement of $Ti^{4+}$ in the anatase $TiO_2$ lattice by cations with an oxidation state larger than +4 can leads to the formation of vacancies at the cation site ($V_{Ti}$) or to formation of interstitial Ti-like defects ($Ti_i$) in which Ti assumes the +3 oxidation state. In both cases independently, doping would inhibit the transition in question [37]. It is important to note that the same element used as a dopant in $TiO_2$, depending on the preparation process, can enter the structure either in a substitutional or interstitial way, and can also assume different oxidation states. Another important factor that directly influences the phase transition is the ionic radius of the dopants, by producing large crystal lattice deformations even in a substitutional doping. In the anatase-rutile transition these deformations must be eliminated or be overcome, in such a way that, from the energetic point of view, the deformations act as an energy barrier stabilizing the anatase phase [30]. Although there is no consensus, in general, dopants with ionic radii larger than that of $Ti^{4+}$ in the anatase structure inhibit the phase transition, while dopants with smaller ionic radii favor the transition [28]. In our case $r_{La} = 1.032$ Å ($N = 6$) [39], thus our results indicates that $La^{3+}$ is in fact doping the anatase $TiO_2$ structure and forming the ternary compound $Ti_{1-x}La_xO_2$.

At this point we can summarize that, to obtain a stable and single phase anatase $TiO_2$ and La-doped anatase $TiO_2$ with relative high crystallinity, the synthesis would be performed at around 500 °C in a $O_2$ atmosphere with gas flux of 2.5 L/min. in our system. $TiO_2$ and La:$TiO_2$ with La nominal concentration of $x_N = 0.06$ were prepared at these conditions. Figure 6(a) presents a representative TEM micrograph for the La:$TiO_2$ sample, it is observed that the powder consists of agglomerations of nanoparticles in the submicron-scale. Figure 6(b) present a high-resolution micrograph showing the {101} planes of the anatase $TiO_2$ (interplanar distance of 0.35 nm). The SAED presented in Figure 6(c) was obtained from the cluster of nanoparticles shown in (a), it reveals a crystalline diffraction pattern corresponding to the anatase $TiO_2$ structure. Several analyses performed in different cluster do not show the presence of any La-related secondary phases.

The powder of the samples was cold compacted uniaxially in the form of pellets for the SEM analyses. Figure 6(d) shows a representative SEM micrograph of the La:$TiO_2$ sample, and the correspondent elemental mapping obtained via EDS. The micrograph was obtained by using a backscattered detector in order to highlight the contrast among any different elemental composition. A series of full scans over large areas over the surface of the sample confirm the absence of La-related secondary phases (clusters of high La content). Both SAED and SEM/EDS results indicate the La doping of the anatase $TiO_2$ structure. The elemental semiquantitative measurement offered by the EDS technique returns an effective La concentration of $x_E = 0.050(2)$, in good agreement with the nominal composition ($x_N = 0.06$). The obtained particle size distribution is presented in Figure (d) and (e) for the $TiO_2$ and La:$TiO_2$ samples, respectively. The statistical results confirm the nanostructured character of the processed powders. With the La addition to the system it is also observed a decrease in the nanoparticle size, as pointed before (Figure 5). The decrease of the size of the nanoparticles is an important observation. It can be associated to the La doping on the surface of the TiO₂ nanoparticles [38, 40, 41]. In the solid-state processing the nucleation of the nanoparticles and its subsequent growth at low temperatures is governed by elemental diffusion processes that strongly depends on to the chemical properties of the surface of the nanoparticles particles [42, 43]. Accordingly, we can infer that the La doping, in a first moment on the surface of the nucleated TiO₂ nanoparticle, decreases its surface energy stabilizing the relatively small nanoparticles [40]. Such similar behavior with adsorbed water on the surface of anatase TiO₂ nanoparticles was reported previously [44].

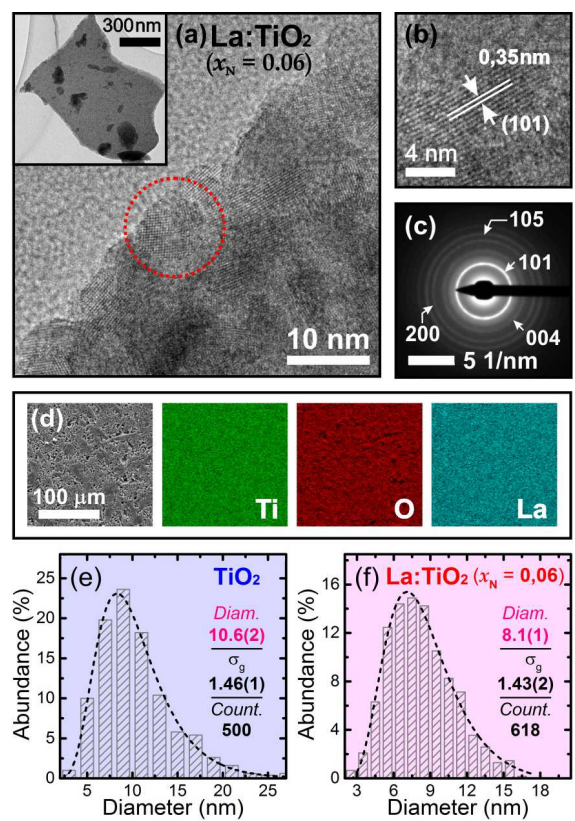

**Figure 6.** (a) Representative TEM micrography of the nanoestructered La:TiO₂ ($x_N$ = 0.06) sample. (b) Correspondent high-resolution image showing the plane lattice correspondent to the family of planes (101). (c) SAED pattern obtained for the aggregated bunch of particles presented in the inset of (a). (d) Representative SEM micrography of the surface of the compacted powder of the La:TiO₂ ($x_N$ = 0.06) sample and corresponding elemental mapping for Ti, O and La. Particle size distribution histogram for the (e) TiO₂, and (f) La:TiO₂ ($x_N$ = 0.06) samples. The dashed line in panel (e) and (f) is the log-normal fit, the correspondent statistical data obtained are also presented in the panels.

Figure 7(a) presents the XRD results obtained for the nanostructured TiO₂ and La:TiO₂ ($x_N$ = 0.06) samples processed at 500 °C and in O₂ atmosphere (2.5 L/min.). It is observed only the diffraction peaks correspondent to the anatase TiO₂ structure (ICSD 9852). For the La:TiO₂ sample the XRD peaks decrease and widen in respect to those for the TiO₂ sample. No secondary phases related to the La were observed within the detection limit of the measurements, indicating that the $La^{3+}$ is effectively doping the anatase $TiO_2$ structure forming the ternary compound $Ti_{0.94}La_{0.06}O_2$. The widening of the XRD peaks for the La:$TiO_2$ sample is associated to the decrease in the nanoparticle diameter, as pointed by the particle size statistical analysis presented above. Figure 7(b) presents the correspondent measured Raman spectra for the studied samples. It is also only observed the vibrational modes related to the anatase $TiO_2$ structure, any secondary phase was detected, confirming the preceding XRD results, and indicating the La doping of the anatase $TiO_2$ structure. It is interesting to call attention to the relative higher intense of the vibrational modes for the La:$TiO_2$ sample. One can find on literature similar experimental reports relating such kind of effect to different sources: structural distortions caused by the La doping due its higher ionic radii [45]; the formation of a core-shell system inducing also structural distortions [46]; and to the presence of $V_O$ mainly on the surface of the studied nanoparticles [47]. Considering that the replacement of $Ti^{4+}$ by $La^{3+}$ favors the formation of $V_O$ due to charge compensation process even under $O_2$-rich conditions, and that the our preceding results indicates that the concentration of $La^{3+}$ is higher on the surface of the nanoparticle, we infer that the observed increase of the Raman modes with the presence of La in our samples it is associated to surface structural defects (structural distortions and/or the formation of $V_O$) due to the $La^{3+}$ doping. Figure 7(c) and 7(d) presents the Tauc plots ($[F(R)h\mathrm{v}]^{1/2}$) and the measured correspondent energy bandgap ($E_g$) for the $TiO_2$ and La:$TiO_2$ samples, respectively. Tauc plots were obtained from reflectance measurements performed in the range of 200-800 nm. There are no significant changes in $E_g$ with the La doping of the anatase $TiO_2$ structure. As stated above, with the La doping it was expected to obtain a reduction in the $E_g$ due to the introduction of $4f$ levels of the La inside the bandgap [14, 48]. However, we have also observed a decrease in the size of the nanoparticles with La doping, which lead us to consider an quantum confinement effect acting in the sense of increasing $E_g$ [49-53]. Therefore, we can attribute the invariance of $E_g$ with the La doping to a compensation process. In spite of the introduction of the $4f$ states inside the $TiO_2$ bandgap decreasing the effective $E_g$, the La doping also leads a decrease of the nanoparticle size leading to an increase of $E_g$ due to a quantum confinement effect.

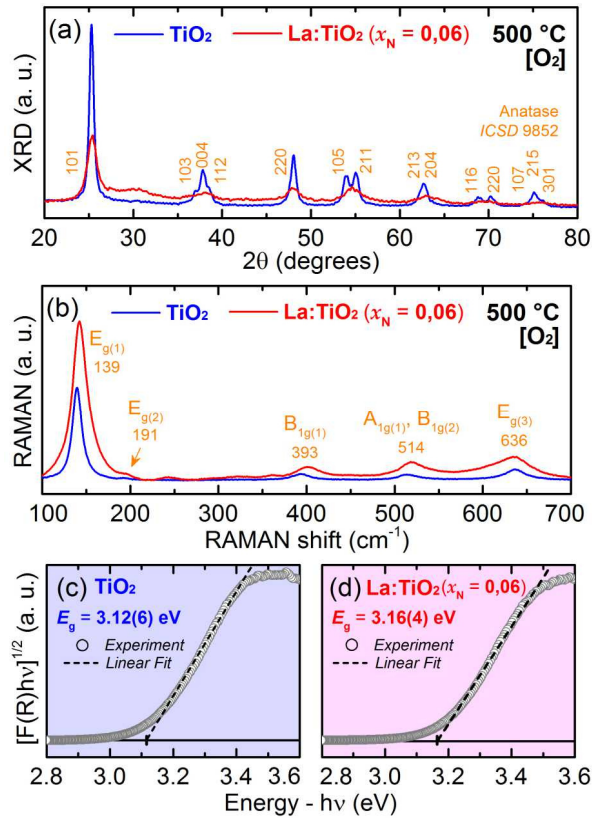

**Figure 7.** (a) XRD diffractograms and (b) Raman spectra of nanoestructered $TiO_2$ and La:$TiO_2$ ($x_N = 0.06$) samples prepared at the temperature of 500 °C in $O_2$ atmosphere (2.5 L/min.). Tauc plots ($[F(R)h\nu]^{1/2}$) as a function of the photon energy ($h\nu$) near the band edge for the (c) $TiO_2$ and (d) $Ti_{0.94}La_{0.06}O_2$ samples.

All the structural analyses presented above gives evidences of the effective La doping of the anatase $TiO_2$ structure. In spite of not observe any change in the $E_g$, as it was desirable, the structural results indicate the La incorporation mainly on the surface of the nanoparticles, what is favorable for its photocatalytic activity. The Raman result indicates the induction of surface structural defects with the La doping, more specifically structural distortions and/or $V_O$. As pointed above, it has been reported that $V_O$ defects are efficient in promoting the separation of photogenerated charge carriers ($e^-/h^+$) increasing the photocatalytic efficiency of anatase $TiO_2$ nanoparticles [19, 20]. Furthermore, several works indicate that donor defects located on the surface of nanoparticles are more efficient in separating charges than defects located in the interior (bulk) of the nanoparticle [21, 22]. For instance, $V_O$ are defects of donor character [54], it means, it introduces electronic levels into the $TiO_2$ bandgap close to conduction band [55]. Once the donor defects are located on the surface of the nanoparticle (Figure 8(a)), there will be a formation of a gradient of potential (internal electrical field) that drives the migration of photogenerated electrons from the bulk to the surface of the nanoparticle (Figure 8(b)), thus increasing the separation of the $e^-/h^+$ pair, decreasing the recombination rate and increasing the photocatalytic efficiency of anatase $TiO_2$ nanoparticle [56, 57]. Another interesting factor regarding the location of defects on the surface of the photocatalytic material is that a higher concentration of defects makes the surface more reactive and, therefore, more interacting with the external environment, which facilitates the adsorption of oxidative/reducing molecules and of molecules to be degraded (pollutants).

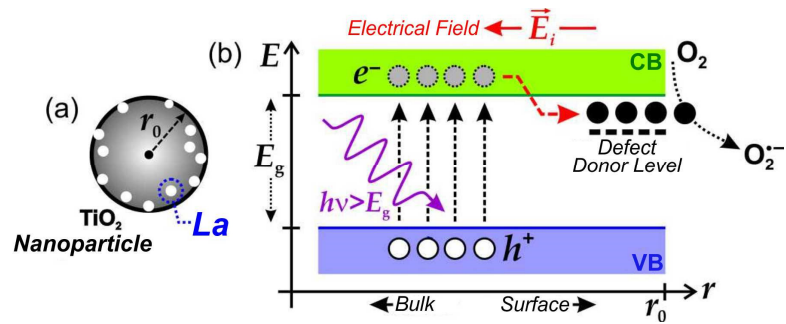

**Figure 8.** (a) $TiO_2$ nanoparticle of radius $r_0$ with $La^{3+}$ predominantly on its surface. (b) Energy band diagram: Energy ($E$) as a function of radial distance ($r$) of the nanoparticle. The structural defects introduce donor levels in the region close to the surface of the nanoparticle. Photogenerated electrons ($e^-$) are driven to the surface of the nanoparticle by the effective internal electric field related to the conduction band profile (as a type-I heterogeneous semiconductor junction), separating the $e^-/h^+$ pair and decreasing the recombination rate, thus increasing the photocatalytic efficiency.

## 4  CONCLUSION

In summary, we have presented a comprehensive study of the main factors related to the processing of the $TiO_2$ and La-doped $TiO_2$ nanopowders in the anatase phase for photocatalytic applications. It was shown that the anatase-rutile transition is promoted by $V_O$, so that the processing of the samples in $O_2$ atmosphere stabilize the anatase phase. The incorporation of La into the $TiO_2$ structure also favors the anatase phase. The doping introduces structural deformations that act as energy barriers to the anatase-rutile transition. Our results show that the suppression of the rutile phase for a specific processing temperature can be used as an indication of an effective doping of the TiO₂ structure. Concerning the processing of the La:TiO₂ samples, the structural characterization confirms the effective doping of the anatase TiO₂ structure forming the ternary compound $Ti_{1-x}La_xO_2$. Due to the higher ionic radii the incorporation of the La into the anatase TiO₂ takes place mainly in the surface of the nanoparticles, decreasing the energy of the surface, and leading to nanoparticles of small sizes. In spite of the La-doping, the $E_g$ remains as the undoped TiO₂ samples, there was no changes of the light absorption spectrum for the La-doped TiO₂ samples. We attributed this behavior to a compensation process associated to a quantum confinement effect due to the decrease of the nanoparticle size with the La-doping. However, the structural results demonstrated the potential formation of $V_O$ due to the La-doping in the surface of the nanoparticles, which would be favorable to the separation of the photogenerated $e^-/h^+$ pair increasing the quantum efficiency of the system and, consequently, increasing its photocatalytic efficiency. The results presented in this report give a valuable contribution to the knowledge of the synthesis of doped nanoparticles for potential applications in photocatalysis.

## ACKNOWLEDGMENTS

Support from agencies CAPES and FAPEMIG is gratefully acknowledged. The authors also acknowledge Dra. M. I. B. Bernardi of the Universidade de São Paulo for the reflectance measurements and Prof. Dr. A. C. Doriguetto coordinator of the Laboratório de Cristalografia of the Universidade Federal de Alfenas were the XRD measurements were performed.

## REFERENCES

1.  Kumar, S.G. and K. Rao, *Comparison of modification strategies towards enhanced charge carrier separation and photocatalytic degradation activity of metal oxide semiconductors (TiO2, WO3 and ZnO).* Applied Surface Science, 2017. **391**: p. 124-148.
2.  Zhang, J., et al., *Photocatalysis: Fundamentals, Materials and Applicatio.* 2018, Singapore: Springer Nature Singapore Pte Ltd.
3.  Wang, A.N., et al., *Diphenylarsinic acid contaminated soil remediation by titanium dioxide (P25) photocatalysis: Degradation pathway, optimization of operating parameters and effects of soil properties.* Science of the Total Environment, 2016. **541**: p. 348-355.
4.  Luttrell, T., et al., *Why is anatase a better photocatalyst than rutile? - Model studies on epitaxial TiO2 films.* Scientific Reports, 2014. **4**.
5.  Cushing, S.K., et al., *Effects of Defects on Photocatalytic Activity of Hydrogen-Treated Titanium Oxide Nanobelts.* Acs Catalysis, 2017. **7**(3): p. 1742-1748.

6.      Zhang, H., et al., *Insights into the effects of surface/bulk defects on photocatalytic hydrogen evolution over TiO2 with exposed {001} facets.* Applied Catalysis B-Environmental, 2018. **220**: p. 126-136.

7.      Borgarello, E., et al., *VISIBLE-LIGHT INDUCED WATER CLEAVAGE IN COLLOIDAL SOLUTIONS OF CHROMIUM-DOPED TITANIUM-DIOXIDE PARTICLES.* Journal of the American Chemical Society, 1982. **104**(11): p. 2996-3002.

8.      Bae, S.W., et al., *Photophysical properties of nanosized metal-doped TiO2 photocatalyst working under visible light.* Journal of the Korean Physical Society, 2007. **51**: p. S22-S26.

9.      Al-Salim, N.I., et al., *Characterisation and activity of sol-gel-prepared TiO2 photocatalysts modified with Ca, Sr or Ba ion additives.* Journal of Materials Chemistry, 2000. **10**(10): p. 2358-2363.

10.     Kang, M., *Synthesis of Fe/TiO2 photocatalyst with nanometer size by solvothermal method and the effect of H2O addition on structural stability and photodecomposition of methanol.* Journal of Molecular Catalysis a-Chemical, 2003. **197**(1-2): p. 173-183.

11.     Litter, M.I. and J.A. Navio, *Photocatalytic properties of iron-doped titania semiconductors.* Journal of Photochemistry and Photobiology a-Chemistry, 1996. **98**(3): p. 171-181.

12.     Yang, Y., et al., *Effect of doping mode on the photocatalytic activities of Mo/TiO2.* Journal of Photochemistry and Photobiology a-Chemistry, 2004. **163**(3): p. 517-522.

13.     Wilke, K. and H.D. Breuer, *The influence of transition metal doping on the physical and photocatalytic properties of titania.* Journal of Photochemistry and Photobiology a-Chemistry, 1999. **121**(1): p. 49-53.

14.     Xu, A.W., Y. Gao, and H.Q. Liu, *The preparation, characterization, and their photocatalytic activities of rare-earth-doped TiO2 nanoparticles.* Journal of Catalysis, 2002. **207**(2): p. 151-157.

15.     Imani, R., et al., *Multifunctional Gadolinium-Doped Mesoporous TiO2 Nanobeads: Photoluminescence, Enhanced Spin Relaxation, and Reactive Oxygen Species Photogeneration, Beneficial for Cancer Diagnosis and Treatment.* Small, 2017. **13**(20).

16.     Yang, J., Y.X. Huang, and J.M.F. Ferreira, *Inhibitory effect of alumina additive on the titania phase transformation of a sol-gel-derived powder.* Journal of Materials Science Letters, 1997. **16**(23): p. 1933-1935.

17.     Chen, C.H., E.M. Kelder, and J. Schoonman, *Electrostatic sol-spray deposition (ESSD) and characterisation of nanostructured TiO2 thin films.* Thin Solid Films, 1999. **342**(1-2): p. 35-41.

18.     Yang, J. and J.M.F. Ferreira, *On the titania phase transition by zirconia additive in a sol-gel-derived powder.* Materials Research Bulletin, 1998. **33**(3): p. 389-394.

19.     Zhang, K.F., et al., *Black N/H-TiO2 Nanoplates with a Flower-Like Hierarchical Architecture for Photocatalytic Hydrogen Evolution.* Chemsuschem, 2016. **9**(19): p. 2841-2848.

20.     Liu, Y.W., et al., *Vacancy Engineering for Tuning Electron and Phonon Structures of Two-Dimensional Materials.* Advanced Energy Materials, 2016. **6**(23).

21.     Yan, J.Q., et al., *Understanding the effect of surface/bulk defects on the photocatalytic activity of TiO2: anatase versus rutile.* Physical Chemistry Chemical Physics, 2013. **15**(26): p. 10978-10988.

22.     Kong, M., et al., *Tuning the Relative Concentration Ratio of Bulk Defects to Surface Defects in TiO2 Nanocrystals Leads to High Photocatalytic Efficiency.* Journal of the American Chemical Society, 2011. **133**(41): p. 16414-16417.

23.     Ranjit, K.T., et al., *Lanthanide oxide-doped titanium dioxide photocatalysts: Novel photocatalysts for the enhanced degradation of p-chlorophenoxyacetic acid.* Environmental Science & Technology, 2001. **35**(7): p. 1544-1549.

24. Li, F.B., et al., *Enhanced photocatalytic activity of Ce3+-TiO2 for 2-mercaptobenzothiazole degradation in aqueous suspension for odour control.* Applied Catalysis a-General, 2005. **285**(1-2): p. 181-189.

25. Doumerc, J.P., et al., *Transition-metal oxides for thermoelectric generation.* Journal of Electronic Materials, 2009. **38**(7): p. 1078-1082.

26. Crisan, M., et al., *Thermal behaviour study of some sol-gel TiO2 based materials.* Journal of Thermal Analysis and Calorimetry, 2008. **92**(1): p. 7-13.

27. Cernea, M., et al., *Structural and thermoluminescence properties of undoped and Fe-doped-TiO2 nanopowders processed by sol-gel method.* Journal of Nanoparticle Research, 2011. **13**(1): p. 77-85.

28. HANAOR, D.A.H. and C.C. SORRELL, **Review of the anatase to rutile phase transformation**. Journal of Materials science, 2011. **46**(4): p. 855-874.

29. ZHANG, H. and J.F. BANFILD, **Thermodynamic analysis of phase stability of nanocrystalline titania**. Journal of Materials Chemistry, 1998. **8**(9): p. 2073-2076.

30. Vargas, S., et al., *Effects of cationic dopants on the phase transition temperature of titania prepared by the sol-gel method.* Journal of Materials Research, 1999. **14**(10): p. 3932-3937.

31. Ihara, T., et al., *Visible-light-active titanium oxide photocatalyst realized by an oxygen-deficient structure and by nitrogen doping.* Applied Catalysis B-Environmental, 2003. **42**(4): p. 403-409.

32. Shannon, R.D. and J.A. Pask, *KINETICS OF ANATASE-RUTILE TRANSFORMATION.* Journal of the American Ceramic Society, 1965. **48**(8): p. 391-&.

33. Mackenzie, K.J.D., *CALCINATION OF TITANIA .4. EFFECT OF ADDITIVES ON ANATASE-RUTILE TRANSFORMATION.* Transactions and Journal of the British Ceramic Society, 1975. **74**(2): p. 29-34.

34. Mackenzie, K.J.D., *CALCINATION OF TITANIA .5. KINETICS AND MECHANISM OF ANATASE-RUTILE TRANSFORMATION IN PRESENCE OF ADDITIVES.* Transactions and Journal of the British Ceramic Society, 1975. **74**(3): p. 77-84.

35. Gamboa, J.A. and D.M. Pasquevich, *EFFECT OF CHLORINE ATMOSPHERE ON THE ANATASE RUTILE TRANSFORMATION.* Journal of the American Ceramic Society, 1992. **75**(11): p. 2934-2938.

36. Syarif, D.G., et al., *Preparation of anatase and rutile thin films by controlling oxygen partial pressure.* Applied Surface Science, 2002. **193**(1-4): p. 287-292.

37. Okada, K., et al., *Effect of silica additive on the anatase-to-rutile phase transition.* Journal of the American Ceramic Society, 2001. **84**(7): p. 1591-1596.

38. de Souza, T.E., et al., *Structural and Magnetic Properties of Dilute Magnetic Oxide Based on Nanostructured Co-Doped Anatase TiO2 (Ti1-xCoxO2-delta).* Journal of Physical Chemistry C, 2013. **117**(25): p. 13252-13260.

39. Shannon, R., *Revised effective ionic radii and systematic studies of interatomic distances in halides and chalcogenides.* Acta Crystallographica Section A, 1976. **32**(5): p. 751-767.

40. da Silva, R.T., et al., *Multifunctional nanostructured Co-doped ZnO: Co spatial distribution and correlated magnetic properties.* Physical Chemistry Chemical Physics, 2018. **20**(30): p. 20257-20269.

41. Mesquita, A., et al., *Dynamics of the incorporation of Co into the wurtzite ZnO matrix and its magnetic properties.* Journal of Alloys and Compounds, 2015. **637**: p. 407-417.

42. Thanh, N.T.K., N. Maclean, and S. Mahiddine, *Mechanisms of Nucleation and Growth of Nanoparticles in Solution.* Chemical Reviews, 2014. **114**(15): p. 7610-7630.

43. Barsoum, M.W., *Fundamentals of ceramics*. 2003, New York ; London: Taylor & Francis.

44. Li, G.S., et al., *High purity anatase TiO2 nanocrystals: Near room-temperature synthesis, grain growth kinetics, and surface hydration chemistry.* Journal of the American Chemical Society, 2005. **127**(24): p. 8659-8666.

45.    Colomban, P. and A. Slodczyk, *Raman intensity: An important tool to study the structure and phase transitions of amorphous/crystalline materials*. Optical Materials, 2009. **31**(12): p. 1759-1763.

46.    Tan, J.Z.Y., F. Xia, and M.M. Maroto-Valer, *Raspberry-Like Microspheres of Core-Shell Cr2O3@TiO2 Nanoparticles for CO2 Photoreduction*. Chemsuschem, 2019. **12**(24): p. 5246-5252.

47.    Lakhera, S.K. and B. Neppolian, *Role of molecular oxygen on the synthesis of Ni(OH)(2)/TiO2 photocatalysts and its effect on solar hydrogen production activity*. International Journal of Hydrogen Energy, 2020. **45**(13): p. 7627-7640.

48.    Stengl, V., S. Bakardjieva, and N. Murafa, *Preparation and photocatalytic activity of rare earth doped TiO2 nanoparticles*. Materials Chemistry and Physics, 2009. **114**(1): p. 217-226.

49.    Serpone, N., D. Lawless, and R. Khairutdinov, *SIZE EFFECTS ON THE PHOTOPHYSICAL PROPERTIES OF COLLOIDAL ANATASE TIO2 PARTICLES - SIZE QUANTIZATION OR DIRECT TRANSITIONS IN THIS INDIRECT SEMICONDUCTOR*. Journal of Physical Chemistry, 1995. **99**(45): p. 16646-16654.

50.    Kormann, C., D.W. Bahnemann, and M.R. Hoffmann, *PREPARATION AND CHARACTERIZATION OF QUANTUM-SIZE TITANIUM-DIOXIDE*. Journal of Physical Chemistry, 1988. **92**(18): p. 5196-5201.

51.    Lee, S., et al., *Correlation of anatase particle size with photocatalytic properties*. Physica Status Solidi a-Applications and Materials Science, 2010. **207**(10): p. 2288-2291.

52.    Xue, X., et al., *Raman Investigation of Nanosized TiO2: Effect of Crystallite Size and Quantum Confinement*. Journal of Physical Chemistry C, 2012. **116**(15): p. 8792-8797.

53.    Knobel, M., et al., *Magnetic and magnetotransport properties of Co thin films on Si*. Physica Status Solidi a-Applied Research, 2001. **187**(1): p. 177-188.

54.    Mamani, N.C., et al., *On the nature of the room temperature ferromagnetism in nanoparticulate co-doped ZnO thin films prepared by EB-PVD*. Journal of Alloys and Compounds, 2017. **695**: p. 2682-2688.

55.    Nowotny, M.K., et al., *Defect chemistry of titanium dioxide. application of defect engineering in processing of TiO2-based photocatalysts*. Journal of Physical Chemistry C, 2008. **112**(14): p. 5275-5300.

56.    Zhang, X.Y., et al., *Surface Defects Enhanced Visible Light Photocatalytic H-2 Production for Zn-Cd-S Solid Solution*. Small, 2016. **12**(6): p. 793-801.

57.    Hou, J.G., et al., *Simultaneously efficient light absorption and charge transport of phosphate and oxygen-vacancy confined in bismuth tungstate atomic layers triggering robust solar CO2 reduction*. Nano Energy, 2017. **32**: p. 359-366.